\documentclass[aps,prl,twocolumn,superscriptaddress,showpacs]{revtex4}
\usepackage{graphicx}

\def\be{\begin{eqnarray}}
\def\ee{\end{eqnarray}}

\begin{document}

\title{Edge solitons of topological insulators and fractionalized
quasiparticles in two dimensions}
\author{Dung-Hai Lee}
\affiliation{Department of Physics, University of California at Berkeley, Berkeley, CA
94720, USA}
\affiliation{Material Science Division, Lawrence Berkeley National Laboratory, Berkeley,
CA 94720, USA}
\author{Guang-Ming Zhang}
\affiliation{Department of Physics, Tsinghua University, Beijing 100084, China}
\author{Tao Xiang}
\affiliation{Institute of Physics, Chinese Academy of Sciences, P. O. Box 603, Beijing
100080, China}
\affiliation{Institute of Theoretical Physics, Chinese Academy of Sciences, P.O. Box
2735, Beijing 100080, China}
\date{\today}

\begin{abstract}
An important characteristic of topological band insulators is the necessary
presence of in-gap edge states on the sample boundary. We utilize this fact
to show that when the boundary is reconnected with a twist, there is always
zero-energy defect states. This provides a natural connection between novel
defects in the two-dimensional $p_x+ip_y$ superconductor, the Kitaev model,
the fractional quantum Hall effect, and the one-dimensional domain wall of
polyacetylene.
\end{abstract}

\maketitle

Excitations carrying fractional quantum numbers (e.g. fractional charge),
such as the quasiparticles in the fractional quantum Hall effect \cite%
{laughlin}, have always been an subject of interest. In 1976 Jackiw and
Rebbi \cite{Jackiw1} wrote a seminal paper which laid the foundation of
charge fractionalization in one spatial dimension. Four years later the
influential paper of Su, Schrieffer and Heeger \cite{Su} proposes the
``Jackiw-Rebbi soliton'' as the charge carrier in doped polyacetylene.
Today, all quantum number fractionalization phenomena, such as the
fractionalization of magnon into spinons \cite{spinon}, can be attributed to
the Jackiw-Rebbi mechanism.

In the last fifteen years, starting with the fractional quantum Hall
effect, condensed matter physicists stumbled upon several instances
where quantum number fractionalization occurs in two spatial
dimensions. These include the quasiparticles of the ``Pfaffian''
quantum Hall state \cite{readmoore}, the
vortices of a spin-polarized $p_{x}+ip_{y}$ superconductor \cite%
{Kopnin,Volovik,Read,Stone,towari}, and the topological excitations in a
spin model proposed by Kitaev \cite{kit}. However, what is lacking is a
general framework, like the Jackiw-Rebbi theory in one dimension, specifying
the condition under which fractionalized excitations will appear. In this
paper we provide such a mechanism and reveal its connection to the
Jackiw-Rebbi theory. In particular, we show that, in two dimensions,
fractional charge will naturally appear around defects of ``topological''
insulators \cite{haldane,kane,qsh}.

Recently it has been shown that the existence of fractional charge in the
quantum Hall effect is connected to the existence of fractionally charged
domain wall in certain one dimensional systems \cite%
{seidel-2005,seidel-2006,karl-2005,karl-2006}. In the following we
generalize such connection and show that when the boundary of a topological
insulator, an insulator which necessarily possesses in-gap edge state, are
reconnected with a twist, there is always zero-energy defect states
possessing fractional quantum number.

Let us begin by considering the Kitaev model \cite{kit}. This exactly
soluble model describes a honeycomb lattice of quantum one-half spins
interacting via three type of nearest-neighbor interactions (Fig.~\ref%
{kitaev}a). The Hamiltonian is given by
\begin{equation}
H={\frac{1}{2}}~\sum_{n\in \mathrm{w}}~\sum_{\mu =1,2,3}~J_{\mu }{\sigma }%
_{n+e_{\mu }}^{\mu }{\sigma }_{n}^{\mu },  \label{kitaev}
\end{equation}%
where ``w/b'' abbreviates for the white/black sublattice (see Fig.~\ref%
{kitaev}a), and $n+e_{\mu }$ is the nearest neighbors of $n$ along
the $\mu $ bond, $\sigma ^{\mu }$ are the Pauli matrices. By
performing Jordan-Wigner transformation it was shown in
Ref.~\cite{xz} that this model is equivalent to a free Majorana
fermion model:
\begin{equation}
H_{M}=-{i}~\sum_{n\in \mathrm{w}}\{~\sum_{\mu =1,2}J_{\mu }\gamma _{n+e_{\mu
}}\gamma _{n}+J_{3}D_{n}\gamma _{n+e_{3}}\gamma _{n}\},  \label{k1}
\end{equation}%
where $D_{n}=\pm 1$ is a classical Ising variable and $\gamma _{n}$'s are
Majorana fermion operators \cite{xz}. Since the honeycomb lattice is
consisted of two sublattices, Eq.~(\ref{k1}) can be recast into
\begin{equation}
H_{M}=\sum_{n\in \mathrm{w},m\in \mathrm{w}}\Psi _{n}^{\dagger }H_{nm}\Psi
_{m},  \label{tt}
\end{equation}%
where $\Psi _{n}^{\dagger }=(\gamma _{n},\gamma _{n+e_{3}})$. In Eq.~(\ref%
{tt}) the $2\times 2$ coupling matrix $H_{nm}$ has value: $H_{nm}=J_{3}D_{n}{%
\sigma }_{2}$ for $n=m$, and $H_{nm}=\mp (iJ_{\mu }/2){\sigma }_{\pm }$ for $%
n\neq m$. The upper sign applies if the bond linking $m$ to $n$ is a
black-to-white bond, and the lower sign applies if it is a white-to-black
bond.

\begin{figure}[tbp]
\centering \includegraphics[scale=0.5]{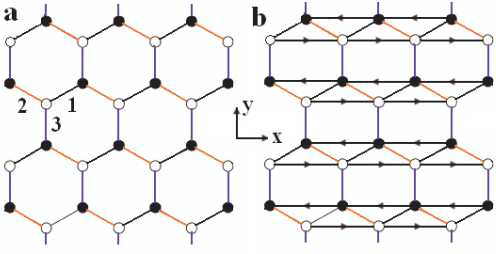} \caption{(a) A
graphical representation of the Kitaev model. There are two
sublattices (white and black), and three types of bonds (labeled
by 1,2,3, or black, red, blue). (b) The graphical representation
of $H+H_{t}$. In the Majorana representation, the three-spin
interactions in Eq.~(\ref{k3}) become next neighbor hopping along
the zigzag chain in the x-direction. The arrow represents the
direction in which the second neighbor hopping matrix elements are
$-iJ_4$.} \label{kitaev}
\end{figure}

In the ground state, the classical Ising variable takes value $D_{n}=1$
modulo a global flip per row \cite{xz}. For $D_{n}=1$, $H_{M}$ is
translation invariant, and can be diagonalized by Fourier transformation.
The Bloch matrix of $H_{M}$ is given by $H_{M}(\mathbf{k})=h_{1}(\mathbf{k}){%
\sigma }_{1}+h_{2}(\mathbf{k}){\sigma }_{2}$ with
\begin{eqnarray}
h_{1}(\mathbf{k}) &=&-J_{2}\sin \alpha (\mathbf{k})+J_{1}\sin \beta (\mathbf{%
\ k}),  \nonumber \\
h_{2}(\mathbf{k}) &=&J_{3}+J_{2}\cos \alpha (\mathbf{k})+J_{1}\cos \beta (%
\mathbf{k}),
\end{eqnarray}%
and $\alpha (\mathbf{k})=(\sqrt{3}k_{x}-3k_{y})/2$, $\beta (\mathbf{k})=(%
\sqrt{3}k_{x}+3k_{y})/2$. It is easy to show that for $J_{1}+J_{2}\geq
J_{3}\geq |J_{1}-J_{2}|$ the energy spectrum is gapless. The possible
connection to the quantum spin liquids has been discussed in Ref. \cite%
{baskaran}. In Ref.~\cite{kit} a magnetic field is introduced to open an
excitation gap in this parameter region, and the non-Abelian quasiparticles
become low energy excitations of this gapped phase.

Unfortunately the magnetic field spoils the integrability of the model. Here
we propose a different way of opening a gap while maintaining the
integrability. This is achieved by adding the following three-spin
interaction to Eq.~(\ref{kitaev})
\begin{equation}
H_{t}={\frac{J_{4}}{2}}\sum_{(ijk)\in \Delta }{\sigma }_{i}^{2}{\sigma }%
_{j}^{3}{\sigma }_{k}^{1}+{\frac{J_{4}}{2}}\sum_{(ijk)\in {\nabla }}{\sigma }%
_{i}^{1}{\sigma }_{j}^{3}{\sigma }_{k}^{2}.  \label{k3}
\end{equation}%
Here $(ijk)$ denote three adjacent sites (with $i$ being the left-most one)
along the zigzag chain running along the x-direction. Depending on whether $%
(ijk)$ form an up-pointing or a down-pointing triangle, we use the first or
second term of Eq.~(\ref{k3}). In terms of the Majorana fermion operators
this amounts to adding a second nearest neighbor hopping between sites along
the zigzag chain:
\begin{equation}
H_{t}=-iJ_{4}(\sum_{i,k\in \mathrm{w}}\gamma _{i}\gamma
_{k}-\sum_{i,k\in \mathrm{b}}\gamma _{i}\gamma _{k}) . \label{snh}
\end{equation}%
Then the Bloch matrix becomes $H_{M}(\mathbf{k})=h_{1}(\mathbf{k}){\sigma }%
_{1}+h_{2}(\mathbf{k}){\sigma }_{2}+h_{3}(\mathbf{k}){\sigma }_{3},$ where $%
h_{3}(\mathbf{k})=2J_{4}\sin (\sqrt{3}k_{x}).$ The vector function $\mathbf{h%
}(\mathbf{k})$ is a continuous mapping from the first Brillouin zone to the
space spanned by $\mathbf{h}=(h_{1},h_{2},h_{3})$. The image is a closed
two-dimensional manifold (henceforth referred as the $h$-surface). Since the
eigenvalues of the Block matrix are $\pm |\mathbf{h}(\mathbf{k})|$, it
follows that if the $h$-surface contains the origin, the spectrum is
gapless, otherwise the spectrum has a gap.

For a $h$-surface not containing the origin, there is an integer topological
index
\begin{equation}
\mathcal{P}=\frac{1}{8\pi }\int d^{2}k~\epsilon ^{\mu \nu }\mathbf{\hat{h}}%
\cdot (\partial _{k_{\mu }}\mathbf{\hat{h}}\times \partial _{k_{\nu }}%
\mathbf{\hat{h}})
\end{equation}%
which counts the number of times the unit vector $\mathbf{\hat{h}}$ wraps
around the origin. As shown in Ref.~\cite{hl}, $\mathcal{P}$ is proportional
to the well known ``TKNN'' index \cite{tknn} in the case of two bands.
Spectra characterized by different $\mathcal{P}$ are topologically distinct.
They can not be deformed into each other without gap closing. In the
parameter regime where non-abelian quasiparticle exists $\mathcal{P}=1$.

\begin{figure}[tbp]
\centering \includegraphics[scale=0.5]{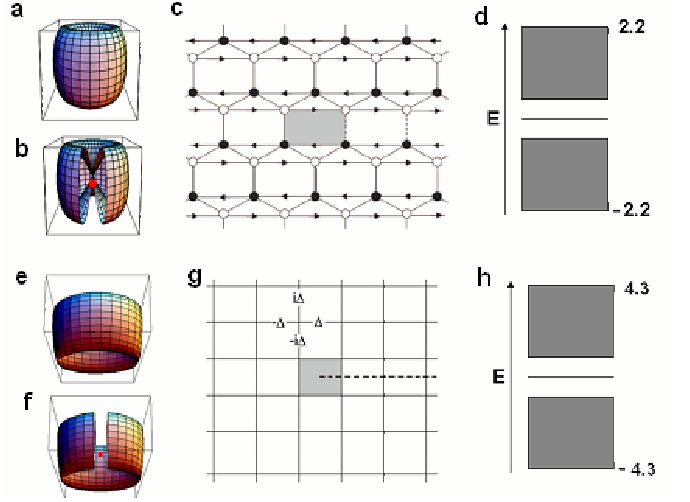} \caption{The
$h$-surface for the Kitaev model (a) and the $p_{x}+ip_{y}$
superconductor (e). In panels (b) and (f) the $h$-surfaces are
dissected to expose the origin (red dot). (c) Topological defects
in the Kitaev model. To the right of the grey-shaded plaquette the
sign of $D_n$'s are reversed. As a result the corresponding
vertical hopping matrix elements change sign. They are shown by
the dashed bonds. (d) The eigenspectrum assoicated with two far
separated topological defects in the Kitaev model with
$J_{1}=J_{2}=1 $, $J_{3}=0.2$ and $J_{4}=0.5$. (g) A vortex
centered at the gray-shaded plaquette in a $p_{x}+ip_{y}$
superconductor. The pairing order parameter is shown for four
bonds. (h) The energy spectrum of two far-separated
vortex-antivortex pair with $t=1,\Delta _{0}=0.5$ and $\protect\mu
=0.3$.} \label{defect}
\end{figure}

Topological excitations of the Kitaev model are created by reversing the
sign of $D_{n}$'s in $H_{M}$ along half a row. This is shown by the dashed
bonds in Fig.~(\ref{defect}c). In Fig.~(\ref{defect}d) we have shown the
result of numerical diagonalization for a system of 1600 sites with toric
boundary condition. Because of the boundary condition two defects are
introduced, they are separated by 100 sites in the x-direction. They
introduce two mid-gap states with a tunnel-splitting (which is already
invisible here) which decreases exponentially with the separation.

Now we switch gear to discuss the vortices in a spin-polarized
$p_{x}+ip_{y}$ superconductor. Let us consider this problem on a
square lattice. Similar to the Kitaev model, the Hamiltonian can
also be written in the form of Eq.~(\ref{tt}), except
$\Psi^\dagger_n=(c^\dagger_n,c_n)$ and $c_n$ is a fermion operator.
The Bloch matrix is characterized by $ \mathbf{h} (\mathbf{k})=
\left( -\Delta_0 \sin{k_y},\, \Delta_0\sin{\ k_x}, \,
-t(\cos{k_x}+\cos{k_y})-\mu \right).$ Here $\Delta _{0}$ is the
pairing amplitude, $t$ is the hopping integral,
and $\mu $ is the chemical potential. The $h$-surface is shown in Fig.~(\ref%
{defect}e) and (\ref{defect}f). Straightforward calculation shows that $%
\mathcal{P}=1$. After a singular gauge transformation, a vortex can be
created by reversing the sign of the hopping matrix elements along a cut as
shown in Fig.~(\ref{defect}g). Explicit calculation shows that there is also
a zero mode associated with each vortex (Fig.~\ref{defect}h).

For each free Majorana or Bogoliubov fermion model discussed above,
there is a free fermion model with identical excitation spectrum. To
obtain this fermion
model we simply replace $\Psi^\dagger_n$ in Eq.~(\ref{tt}) by the fermion operator $%
\Psi^\dagger_n=(c^\dagger_{1n},c^\dagger_{2n})$ where $1$ and $2$
are ``flavor'' indices (they might represent the two sites in the
unit cell of a lattice). 
This fermion model acts as a representative of all models which
share the same $H_{nm}$ (hence the same eigen spectrum). However,
while the representative fermion model is global U(1) invariant, the
Majorana and Bogoliubov fermion models only have $Z_{2}$ symmetry.
The fact that in the $Z_{2}$ models the particle number is only
conserved modulo two is the root of non-abelian statistics. In the
rest of the discussion, we refer to the system described by a gapped
free fermion model with non-zero $\mathcal{P}$ as a ``topological
band insulator''. Thus the representative fermion models for the
Kitaev model and the $p_{x}+ip_{y}$ superconductor are topological
band insulators. Knowing the properties of edge states and defects
of the representative fermion model, one can readily deduce the
corresponding properties of the Majorana fermion (Kitaev) or the
Bogoliubov fermion ($p_{x}+ip_{y}$) models with the same $H_{nm}$.
For example, while in the fermion model the edge states are free
fermions, and the defect zero modes carry half fermion quantum
numbers, those in the Majorana/Bogoliubov fermion model are true
Majorana fermions.

In the following we provide an unifying mechanism for the appearance
of defect zero mode when the representative fermion model describes
a topological band insulator.
\begin{figure}[tbp]
\centering \includegraphics[scale=0.55]{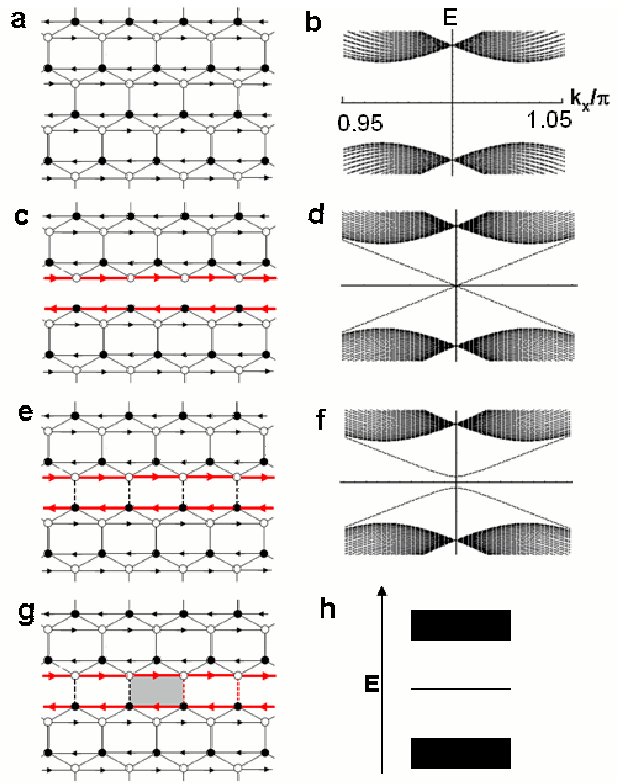} \caption{ (a) The
Kitaev model and (b) its eigen-spectrum as a function of wavevector
$k_x$ along the x-direction around $k_x=\protect\pi $. (c) One row
of bonds along $\hat{x}$ are removed and (d) the corresponding
energy spectrum. The arrowed-red lines in (c) indicate the edge
states are chiral. (e) A weaker hopping between the edge  (dashed
black vertical bonds) is reintroduced. (f) The gapped energy
spectrum corresponds to (e). (g) An edge soliton (marked by the gray
plaquette) is introduced by reversing the sign for half of the
vertical bonds (red dashed lines) between the edges. (h) The energy
spectrum corresponds to (f). The spectra in (b), (d) and (h) are
obtained from $H_{F}$ with $J_{1}=J_{2}=1$, $J_{3}=0.2$ and
$J_{4}=0.5$. The spectrum in (f) and (h) was obtained with a
restored edge coupling $J_{3}=\pm 0.02$.} \label{proj}
\end{figure}
As an example, let us consider the fermion representative of the
Kitaev model (Fig.~\ref{proj}a). Fig.~(\ref{proj}b) shows the gapped
spectrum as a function of momentum along the longitudinal circle.
Due to its topological nature, if we remove a row of bonds
(Fig.~\ref{proj}c), in-gap edge states appear\cite{hastugai} as
shown  in Fig.~(\ref{proj}d). The left and right moving chiral edge
fermions, represented by the arrowed red lines in
Fig.~(\ref{proj}c), are described by a massless Dirac Hamiltonian in
1D. If we reconnect the two edges, but with weaker bonds, a smaller
gap reappear in the edge spectrum (see Fig.~\ref{proj}e and
Fig.~\ref{proj}f). The edge fermions are now described by a massive
Dirac Hamiltonian
\begin{equation}
H_{E}=\int dx~(-iv\psi ^{\dagger }{\sigma }_{z}\partial _{x}\psi +m\psi
^{\dagger }{\sigma }_{x}\psi ),  \label{dirac1}
\end{equation}%
where $v$ is the edge velocity, $\psi^\dagger
=(\psi_{R}^\dagger,\psi _{L}^\dagger)$ with $\psi _{R/L}^\dagger$
being the right/left fermion creation operators. When the restored
bonds have a sign reversal (Fig.~\ref{proj}g), the mass term in
Eq.~(\ref{dirac1}) becomes x-dependent and changes sign as $x$ goes
through the location of the topological defect. This should result
in one localized zero mode per defect according to
Ref.\cite{Jackiw1,Su}. Fig.~(\ref{proj}h) shows that this is indeed
true. The presence of such zero mode plus the fact that the spectra
of Majorana fermion models are $E\leftrightarrow -E$ symmetric,
immediately imply that the zero modes are Majorana fermion states in
the original Kitaev model. Furthermore, using the argument of
Ref.\cite{Ivanov} it can be shown that the braiding of such Majorana
fermion defects leads to the non-Abelian statistics. Thus, via the
mechanism of Refs.\cite{Jackiw1,Su} a \textit{two dimensional}
defect with fractionalized quantum number has emerged! Its presence
is determined by the topological nature of the host bulk band
insulator just as the edge states are.

Actually, a similar phenomenon was also found in the continuum
theory of Dirac fermions interacting with the topological defects of
a Higgs field\cite{ch}. However, in all the examples we considered
here, the location of the \textit{edge} Dirac point in the momentum
space is far away from those of the \textit{bulk} Dirac points.
Consequently the theory discussed in Ref.~%
\cite{ch} is not applicable here.

\begin{figure}[tbp]
\includegraphics[scale=0.5]{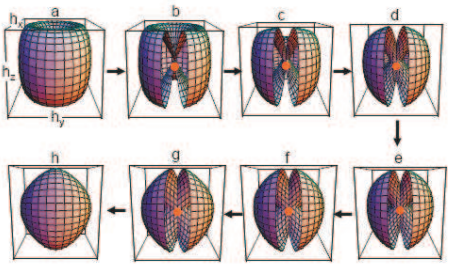}
\caption{The evolution of the of the $h$-surface as the model in Fig.~(\ref%
{kitaev}b) is gradually deformed into Haldane's model. In constructing the
figure we used $J_{1}=J_{2}=J_{3}=1$. The full strength of the second
neighbor hopping is $0.5$.}
\label{topoevol}
\end{figure}

As a digression, we now show the fermion representative of the
Kitaev model is topologically equivalent to Haldane's lattice model
for integer quantum Hall effect\cite{haldane}. Fig.~(\ref{topoevol})
shows the evolution of the $h$-surface by gradually switching off
the second neighbor hopping  in Haldane's model which are not
contained in Eq.~(\ref{snh}). The leftmost column are the
$h$-surfaces for the Haldane model (top) and Kitaev (bottom),
respectively. In the rest of the figure the surfaces are dissected
to reveal the origin (the red dot). Clearly as we follow the
evolution the origin never migrate across the $h$-surface. Thus
$\mathcal{P}$ for the two models are the same.

Finally what about the Laughlin quasiparticles? Although fractional
quantum Hall liquids are not band insulators they are clearly {\it
topological insulators}. Indeed, as shown by Wen \cite{wen2}, when a
quantum Hall liquid on a torus is cut open (Fig.~\ref{proj}a), there
are ``chiral Luttinger liquid'' in-gap edge modes. At $1/m$ filling,
the edge modes are described by the following free boson Hamiltonian
\begin{equation}
H_{B}=\int dx\Big\{{\frac{2\pi }{m}}\Pi (x)^{2}+{\frac{m}{8\pi }}[\partial
_{x}\phi (x)]^{2}\Big\},  \label{boson}
\end{equation}%
where $\Pi $ and $\phi $ are conjugate boson fields satisfying $[\Pi
(x),\phi (y)]=i\delta (x-y)$. To reconnect the edges, a potential $V=-g\int
dx\cos (m\phi )$ needs to be added \cite{wen2}. When $g$ is sufficiently
big, a gap opens in the edge spectrum and the ground states become $m$%
-fold degenerate. They are characterized by $\langle \phi (x)\rangle
=2\pi l/m\,(l=0,...,m-1)$. In this case an edge soliton is where
$\phi (x)$ interpolates between two different ground states, say,
$\langle \phi (x)\rangle =0$ and $2\pi /m$. As shown by Goldstone
and Wilczek \cite{gw}, such a soliton carries a charge $\Delta
Q=\Delta \phi /2\pi =1/m$, precisely the same as a that of a
Laughlin quasiparticle \cite{laughlin}. Hence the Laughlin
quasiparticles are also edge solitons of a topological insulator!

\textrm{Acknowledgement} DHL was supported by DOE Contract No.
DE-AC02-05CH11231. GMZ and TX acknowledge the support from the NSF-China and
the national program for basic research.

\end{document}